# AN EXTENSION OF WEAK-VALUE THEORY FOR BIREFRINGENCE-INDUCED DISPLACEMENT MEASUREMENTS

Garrett Josemans, Benjamin Baldwin, John E. Gray, Patrick Graves, and Kevin Bertschinger

*Naval Surface Warfare Center, Dahlgren Division, Electromagnetic and Sensor Systems Department*

We present an investigation into the tilt sensitivity of the canonical, optical, weak-value amplification device (COWVAD) − introduced by Duck, Stevenson, and Sudarshan in 1989 – for potential application in a Coriolis vibratory gyroscope (CVG). We model the breakdown of the weak-value amplification effect in this device with respect to angle of incidence between the laser beam and the surface of the birefringent crystal. Additionally, we model the effect of beam divergence due to misalignment of the crystal's optic axis. We found that the presence of beam divergence allows the architecture to be placed in an inverse weak-value amplification configuration. We found that weak-value amplification occurs at angles other than those reported in Duck 1989. We also present laboratory measurements that support the validity of the mathematical model. The maximum tilt sensitivity measured was ∼.58 m/rad, which, while 390 times greater than it would be without weak-value amplification, does not represent the maximum possible sensitivity of the device through weak value amplification.

## I. INTRODUCTION

Weak values, since their introduction in 1988 by Aharonov et. al. [1], have made an impressive expansion into many different areas of research including optical interferometry [2,3,4,5,6], gravimetry [7], signal processing [8], and, very recently, rotation rate measurements [9] and fully atomic weak values [10]. Weak values have also proved useful in uncovering some very strange phenomenon including retrocausality [11] and non-local effects [12]. The most prominent feature of weak-value theory is the amplification of small effects, referred to as weak-value amplification (WVA), which has been applied to measuring very small longitudinal displacements [13], changes in gravity [7], time delays [14], and magnetic fields [15]. In each case the theory of weak values has been molded to fit the application, but in doing so there are aspects of the physical behavior of the system that are often over-looked or controlled-for in order to allow a simpler description and theoretical model. Such is the case with one of the original examples of WVA, the measurement of the displacement or divergence of two orthogonally polarized beams of light caused by a birefringent crystal [16,17,18,19].

Each of the previous theoretical treatments lack significant generality to apply WVA to a broad range of applications wherein the control provided by laboratory experiments is traded for the barrage of variables characteristic of real world environments. Traditionally this transition is considered well within the engineering arena but sometimes theoretical developments must be made to account for phenomenon that arise. Such is the case for the canonical birefringent example. In this report, we take a deep dive into the traditional weak-value theory used to describe the canonical, optical, weak-value amplification device (COWVAD) and expand the interaction unitary to capture the full dependence of the incidence angle $\theta$ of the beam with the surface of the birefringent crystal. This includes the polarization-independent spatial translation $\gamma_o$ of the photons relative to the initial photon path as well as the relative phase-difference $\phi$ between the ordinary and extraordinary rays after exiting the birefringent crystal. The expanded unitary also includes a momentum boost term that accounts for a potential small divergence of the two beams after exiting the crystal. Along with these advancements, we evaluate the breakdown of traditional weak-value theory and the resulting emergent behavior. As an interesting side note, we reveal the presence of WVA at incidence angles heretofore unaddressed by the COWVAD literature in Appendix B.

## II. COWVAD ARCHITECTURE

The COWVAD architecture itself is straightforward, and its utility as an example of WVA is well understood. FIG. 1 illustrates the basic components of the COWVAD consisting of a laser, preselection polarizer, birefringent crystal with optical axis oriented along the x-axis, post-selection polarizer, and an optical detector.

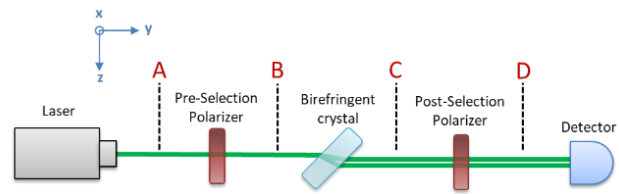

FIG. 1 The Canonical, Optical, Weak-Value Amplification Device. At point "A", the laser beam propagating along the y-axis has arbitrary polarization and an approximately Gaussian intensity profile. At point "B", the beam has passed through a linear polarizer, preselecting the beam into an equal superposition of horizontal (z-axis) and vertical (x-axis) polarization states. At point "C", the birefringent crystal has weakly interacted with the beam, spatially translating each polarization component by a slightly different amount via refraction. The translated beams are largely overlapping so that the two beams are spatially indistinguishable. At point "D," the beam has been post-selected by the second polarizer into a polarization state that is nearly orthogonal to the preselected state.

This prescription will result in the observation of a spatial translation of the Gaussian beam profile (called the "pointer") that is much larger than expected from refraction if key conditions are met. First, the spatial separation between the two beams must be very small (the quantitative description of this distance is left for the Theory section). Second, the angle between the face of the crystal and the incident laser beam must be such that the two beams emerge in phase with one another. Depending on the thickness of the crystal and the wavelength of the laser, there are generally multiple incidence angles where this occurs; these angles are called coherency points. Though the behavior of the COWVAD's pointer at coherency points is well understood, its behavior in the transition region between where WVA does and does not occur, i.e., small angular deviations around coherency points, has not been explored. Experimental observations indicate that, in this region, the pointer rapidly shifts from its amplified to its unamplified position in a manner that is highly sensitive to incidence angle. This previously unreported behavior implies the utility of the COWVAD as the tilt sensitive component in a practical sensor.

### III. THE COWVAD AS A CVG

CVGs are gyroscopes wherein a small vibrating structure responds to rotation about a particular axis due to the Coriolis Effect. A very basic architecture for a potential COWVAD CVG is illustrated in FIG. 2. The proposed device operates by driving a crystal to oscillate at a set frequency $\omega$ and amplitude $X$ along or about a particular axis called the drive axis, $D$. When the whole device is rotated at a rate of $\Omega$ about an axis orthogonal to $D$, let us call this axis $R$, a vibration is induced in the third remaining orthogonal axis, $S$. The amplitude $Y$ of the induced oscillation is proportional to the drive signal parameters and the rotation rate about the $R$ axis, $Y \propto \Omega \omega X$. Thus, $\Omega$ can be determined from measurement of $Y$. The role of the COWVAD is to amplify the measurement of $Y$, thus increasing the sensitivity of the CVG to the rotation rate $\Omega$. The WVA component of the COWVAD requires the crystal's incidence angle $\theta$ to be initialized to a coherency point $\theta_{CP}$, and the polarization angles of the pre- and post-selection polarizers to be chosen appropriately. In this configuration, a Coriolis-Effect-induced oscillation of $\theta$ about $\theta_{CP}$ will result in an oscillating spatial translation of the pointer, amplified by WVA.

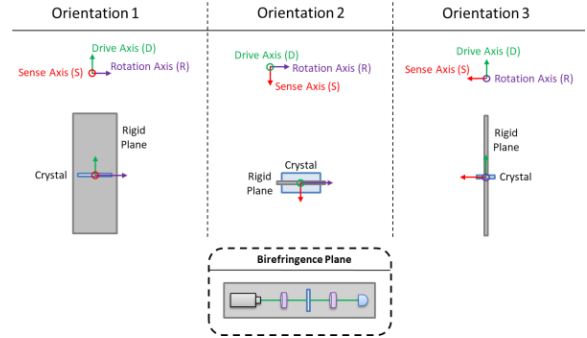

FIG. 2   Simple CVG conceptual model.

The various components associated with the WVA readout must be mounted on a rigid plane that we refer to as the birefringence plane (BP). The crystal must be free to oscillate independently of the BP about D and S, but any rotation of the crystal about R must occur in unison with all other components rigidly mounted to the BP. Conveniently, the COWVAD architecture has relatively minimal components and balancing requirements and, therefore, lends itself well to compactness. There are still some engineering questions to be answered in regards to how small the COWVAD CVG can be made, but the architecture is promising.

### IV. INCIDENCE ANGLE DEPENDENCE

The central feature of the COWVAD is a uniaxial, birefringent crystal oriented with its optic axis perpendicular to the plane of incidence. In this orientation, the crystal has two important effects on the laser beam. First, it acts as a type of polarizing beam-splitter. Light refracting through the crystal follows Snell's Law, but with different indices of refraction depending on polarization. The incident laser beam is prepared in a state of linear polarization oriented $\pi/4$ rad from the crystal's optic axis (x-axis). In this state, the laser beam is split 50-50 along two possible paths, one corresponding to "vertical" polarization (parallel to the crystal's optic axis) and the other corresponding to "horizontal" polarization. The beams exit the crystal propagating along paths parallel to the incident beam but having been spatially translated along the z-axis (FIG. 3).

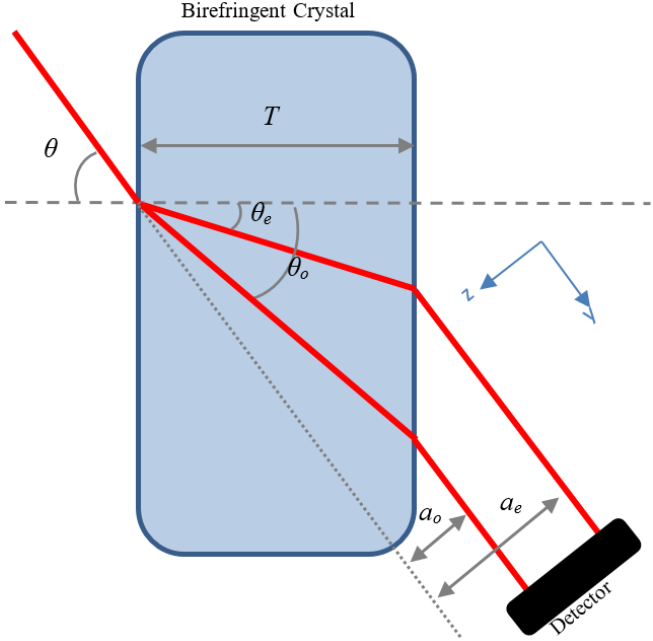

FIG. 3 Refraction through the birefringent crystal. The separation distance between the beams is greatly exaggerated in the diagram, as typically the beams are spatially unresolvable. Note that the optic axis of the crystal is oriented out of the page, perpendicular to the plane of incidence.

The magnitudes of the spatial translations, $a_o$ for the ordinary (horizontally polarized) ray and $a_e$ for the extraordinary (vertically polarized) ray, are dependent on the incidence angle $\theta$ between the laser beam and the crystal face. The $\theta$ dependence of $a_o$ and $a_e$ is given by

$$a_{o,e} = T \sin\theta \left( \frac{n_{air} \cos\theta}{\sqrt{n_{o,e}^2 - (n_{air} \sin\theta)^2}} - 1 \right) \quad (1)$$

where $T$ is the thickness of the crystal and $n_o$, $n_e$, and $n_{air}$ are the relevant indices of refraction. The second important effect that the crystal has on the laser beam is that it acts as a phase retarder. Due to the difference in the optical path lengths traveled by each beam, a relative phase-shift $\phi$ is introduced between them. The $\theta$ dependence of $\phi$ is given by

$$\phi = T k_o \left( \sqrt{n_e^2 - (n_{air} \sin\theta)^2} - \sqrt{n_o^2 - (n_{air} \sin\theta)^2} \right) \quad (2)$$

where $k_o$ is the vacuum wavenumber of the laser. Coherency points are incidence angles for which $\cos\phi = 1$, i.e., the relative phase-shift is effectively zero.

## V. QUANTUM DESCRIPTION AND THE EFFECT OF POST-SELECTION

An ensemble of photons is refracted through the birefringent crystal, entangling the polarization states (qubit) with the transverse z-coordinates (pointer). The ensemble is preselected into an equal superposition of orthogonal polarization states such that the initial state of the qubit takes the form

$$|\psi_i\rangle = \frac{1}{\sqrt{2}}(|H\rangle + |V\rangle) \quad (3)$$

where $|H\rangle$ and $|V\rangle$ are the horizontal and vertical polarization basis states, respectively. The initial pointer state has a distribution that is approximately Gaussian, and thus takes the form

$$|\Phi_i\rangle = \int_{-\infty}^{\infty} dz \, \langle z|\Phi_i\rangle |z\rangle \; : \; \langle z|\Phi_i\rangle = \frac{e^{-z^2/4\sigma^2}}{\sqrt[4]{2\pi\sigma^2}} \quad (4)$$

where $2\sigma$ is the $e^{-2}$ radius of the photon beam and $z$ is the direction perpendicular to photon propagation and parallel to the horizontal, as shown in FIG. 1. The combined state of the qubit and pointer is $|\psi_i\rangle \otimes |\Phi_i\rangle$. As was done in ref. [16], the crystal interaction is modeled as a relative spatial translation $\pm \gamma$ between the $|H\rangle$ and $|V\rangle$ paths. In order to capture the full $\theta$ dependence, the polarization-independent spatial translation $\gamma_o$ of the photons relative to the initial photon path is accounted for, as well as the relative phase-shift $\phi$ between the two paths. In addition, if the optic axis of the crystal is not perfectly aligned with the x-axis, the photon paths exiting the crystal will not be parallel. This is modeled as a relative momentum boost of strength $k$. Together these effects are modeled by the unitary operator

$$\hat{U} = e^{-\frac{i}{\hbar}(\gamma_o + \gamma \otimes \hat{A})\hat{p}} e^{\frac{ik}{\hbar 2} \hat{z} \otimes \hat{A}} e^{-i\frac{\phi}{2} \hat{A}} \quad (5)$$

where $\hat{A} = |H\rangle\langle H| - |V\rangle\langle V|$ is the $\sigma_3$ Pauli spin operator operating on $|\psi_i\rangle$, $\hat{p}$ is the momentum operator operating on $|\Phi_i\rangle$, and $\hat{z}$ is the position operator operating on $|\Phi_i\rangle$. The spatial translation magnitudes of the $|H\rangle$ and $|V\rangle$ paths in (1) are related to the interaction strength variables in (5) by $a_o = \gamma_o - \gamma$ and $a_e = \gamma_o + \gamma$, which yields

$$\gamma_o = \frac{T \cos\theta \, n_{air} \sin\theta}{2} \left( \frac{1}{\sqrt{n_o^2 - (n_{air} \sin\theta)^2}} + \frac{1}{\sqrt{n_e^2 - (n_{air} \sin\theta)^2}} - 2 \right) \quad (6)$$

and

$$\gamma = \frac{T \cos\theta \, n_{air} \sin\theta}{2} \left( \frac{1}{\sqrt{n_o^2 - (n_{air} \sin\theta)^2}} - \frac{1}{\sqrt{n_e^2 - (n_{air} \sin\theta)^2}} \right). \quad (7)$$

After interaction with the crystal, the photon is post-selected via projection onto a final state that is nearly orthogonal to the initial preselected state

$$|\psi_f\rangle = \cos\beta \, |H\rangle + \sin\beta \, |V\rangle \quad (8)$$

where $\beta$ is the angle between the transmission axis of the post-selection polarizer and the z-axis. Combining (3), (4), (5), and (8) yields the final state of the pointer

$$|\Phi_f\rangle = \langle\psi_f|\hat{U}|\psi_i\rangle =$$
$$\frac{1}{\sqrt{2}}\left(\cos\beta\, e^{-\frac{i}{\hbar}\gamma\hat{p}}e^{\frac{ik}{\hbar 2}\hat{z}} + \sin\beta\, e^{\frac{i}{\hbar}\gamma\hat{p}}e^{-\frac{ik}{\hbar 2}\hat{z}}e^{-i\phi}\right)\times \quad (9)$$
$$e^{i\frac{\phi}{2}}e^{-\frac{i}{\hbar}\gamma_o\hat{p}}|\Phi_i\rangle$$

where $\gamma_o$, $\gamma$, and $\phi$ are functions of $\theta$ given by (6), (7), and (2). While the momentum boost parameter $k$ also carries some $\theta$-dependence, it is tied to the misalignment between the crystal's optic axis and the z-axis. For this reason, the $\theta$-dependence of $k$ is not treated explicitly.

From (9), the full expression for the position expectation value of the final pointer state is

$$\langle z\rangle = \langle\Phi_f|\hat{z}|\Phi_f\rangle = \gamma_o + \sigma\times$$
$$\left(\frac{\frac{\gamma}{\sigma}\sin(2\epsilon) + k\sigma\cos(2\epsilon)\sin\phi\, e^{-\frac{(k\sigma)^2+(\gamma/\sigma)^2}{2}}}{1-\cos(2\epsilon)\cos\phi\, e^{-\frac{(k\sigma)^2+(\gamma/\sigma)^2}{2}}}\right). \quad (10)$$

To compare with weak-value theory, consider the case that $k\sigma = 0$ (no momentum boost) and that $e^{-\gamma^2/2\sigma^2}\approx 1$ (weak interaction in $\gamma$ meeting the conditions in (A4)),

$$\langle z\rangle = \gamma_o + \gamma\frac{\sin(2\epsilon)}{1-\cos(2\epsilon)\cos\phi\, e^{-\frac{(k\sigma)^2+\left(\frac{\gamma}{\sigma}\right)^2}{2}}} \quad (11)$$
$$\approx \gamma_o + \gamma\cot\epsilon$$

where the approximate form is valid when $\cos\phi\approx 1$ (coherency points) and the remaining $\epsilon$ terms reduce by the relation $\sin(2\epsilon)[1-\cos(2\epsilon)]^{-1} = \cot\epsilon$. This matches the weak-value theory prediction from [16] and (A5). Alternatively, consider the presence of a weak momentum boost in the inverse-WVA limit, i.e., when $|A_w|\gg 1$ or, equivalently, when $\epsilon\ll k\sigma$ [6]. In this limit, the second term dominates eqn. (10), which reduces to

$$\langle z\rangle = \frac{k\sigma^2\sin\phi}{e^{\frac{((k\sigma)^2+(\gamma/\sigma)^2)}{2}} - \cos\phi}$$
$$\approx \frac{2\phi}{k\left[1+\left(\frac{\gamma}{k\sigma^2}\right)^2\right]} \quad (12)$$

where the approximate form is valid for $\phi\ll 1$. This form matches the inverse-WVA curve derived in eqn. (1) of ref. [2] for a modified Sagnac architecture, which produces a weak momentum boost ($k\sigma\ll 1$) but no spatial translation ($\gamma/\sigma = 0$). The COWVAD always produces a non-zero $\gamma$-value for non-zero incidence angles; nevertheless (12) demonstrates that in this regime the pointer sensitivity to crystal tilt angle (via phase-shift) is inversely proportional to the strength of the momentum boost.

The relative phase-shift $\phi$ generally goes through multiple cycles over the full range of incidence angles, the number of which increases in proportion to crystal thickness. After post-selection, the photon probability density at coherency points is heavily attenuated (FIG. 4).

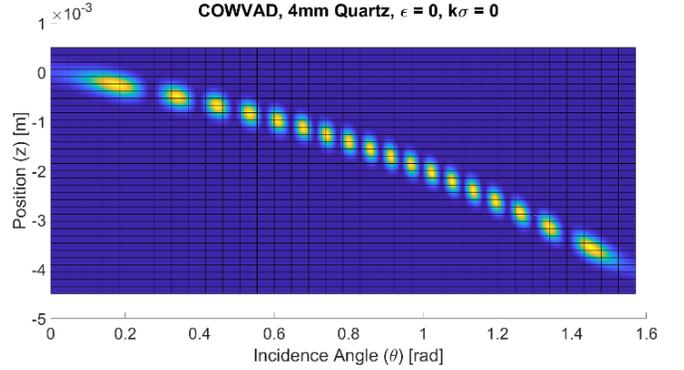

FIG. 4: Position probability density $|\Phi_f|^2$ dependence on incidence angle. The downward trend demonstrates the $\theta$ dependence of $\gamma_0$. The dark vertical bands are the highly attenuated coherency points where $\phi$ approaches integer multiples of $2\pi$ rad.

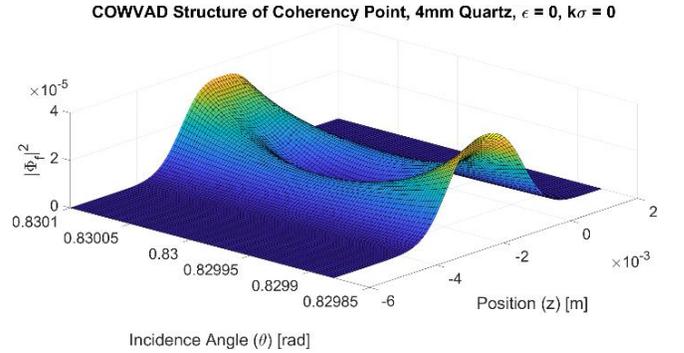

FIG. 5  Probability density $|\Phi_f|^2$ structure at 8th coherency point with crossed polarizers ($\epsilon = 0$) and no momentum boost ($k\sigma = 0$). The interesting behavior lies in the structure of the photon probability density at and near the coherency points. At these points, perfectly crossed pre- and post-selection polarizers produce a symmetric, dual-mode Gaussian distribution. The introduction of a post-selection state such that $\gamma/\sigma \ll \epsilon \ll 1$, or a momentum kick such that $\epsilon \ll k\sigma \ll 1$, breaks the symmetry causing $\langle z\rangle$ to demonstrate WVA or inverse-WVA characteristics, respectively.

## VI. BREAKDOWN OF WEAK-VALUE CONDITIONS

When the weak-value conditions in Eqn. (A4) are not met, the validity of weak-value theory breaks down. The exact structure of this breakdown, specifically with respect to the incidence angle $\theta$ of the birefringent crystal, is the main

## A. Breakdown of Amplification Due to Post-Selection

First consider the behavior at coherency points when the weak-value conditions in Eqn. (A4) are violated. These conditions put rough upper and lower bounds on $\epsilon$. Because the interaction strength $\gamma$ is a function of $\theta$, the rough lower bound of $\epsilon$ is determined at each coherency point. This also puts an upper bound on the WVA that can be achieved, $A_w = \cot\epsilon \ll \cot(\gamma/\sigma)$. When the weak-value conditions for $\epsilon$ are met, the resulting beam profile is a single-mode Gaussian. In this regime, the position expectation value $\langle z \rangle$ is virtually identical to the peak of the Gaussian. For very small $\epsilon$ values ($\epsilon \lesssim \gamma/\sigma$), a secondary mode emerges in the beam profile, growing in magnitude until a symmetric dual-mode distribution occurs at $\epsilon = 0$ (FIG. 5). As the magnitude of the second mode approaches that of the first, $\langle z \rangle$ decouples from the weak-value predictions, as shown in FIG. 6. Though not well approximated by weak-value theory, the largest shift in position expectation occurs when $\epsilon = \gamma/2\sigma$. Note that when $\epsilon \ll \gamma/2\sigma$, the COWVAD device enters the inverse-WVA regime, where it becomes sensitive to small changes in $\epsilon$, as shown by the steep slopes near the y-axis in FIG. 6. For larger values of $\epsilon$, the amplification diminishes and $\langle z \rangle$ approaches $\gamma_o$ as $\epsilon$ approaches $\pi/4$. Note from (10) that the presence of a weak momentum boost ($k\sigma \sim \gamma/\sigma$) has very little impact on $\langle z \rangle$ exactly at coherency points due to the fact that a zero phase-shift zeros-out the momentum boost term in the numerator.

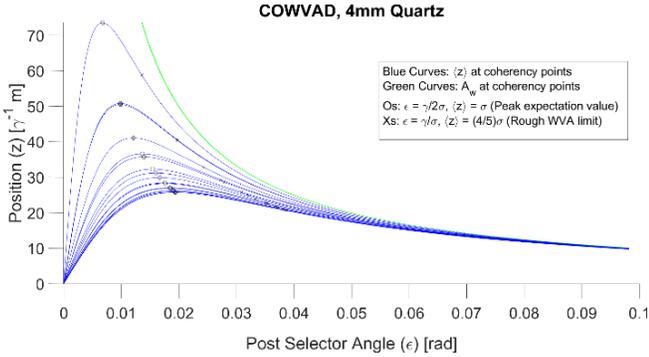

FIG. 6 Position expectation at the coherency points. The position expectation at coherency points $\langle z \rangle$ (blue) is well approximated by weak-value theory (green) when $\gamma/\sigma \ll \epsilon \ll 1$. The largest values of $\langle z \rangle$ occur when $\epsilon = \gamma/2\sigma$ (black circles). Note that at $\epsilon = \pi/4$ (not pictured), the blue and green curves cross 1, meaning $\langle z \rangle = \gamma$, or no weak-value amplification occurs. Note that $\langle z \rangle = \sigma$ at all peaks.

## B. Breakdown of Amplification Due to Incidence Angle

Next, consider how amplification breaks down at a fixed $\epsilon$ due to the deviation of $\theta$ away from a coherency point. The photon probability distribution is dependent on $\theta$ as seen in FIG. 4, FIG. 5, FIG. 7, and FIG. 11. Therefore, deviations in $\theta$ result in large intensity changes. In the case of zero momentum boost, as $\theta$ (and therefore $\phi$) deviates from the coherency point, $\langle z \rangle$ rapidly shifts from the value predicted by weak-value theory, $\gamma_o + A_w\gamma$, to the more intuitive, classical value of $\gamma_o$, as shown in FIG. 7. The slope of the pointer shift, i.e., the sensitivity of $\langle z \rangle$ to $\theta$, is determined mainly by $\gamma$, $\epsilon$, and $\sigma$ (which control the magnitude of the shift), and the thickness of the crystal $T$ and wavenumber of the photon $k_o$ (which determine the width of the shift with respect to $\theta$).

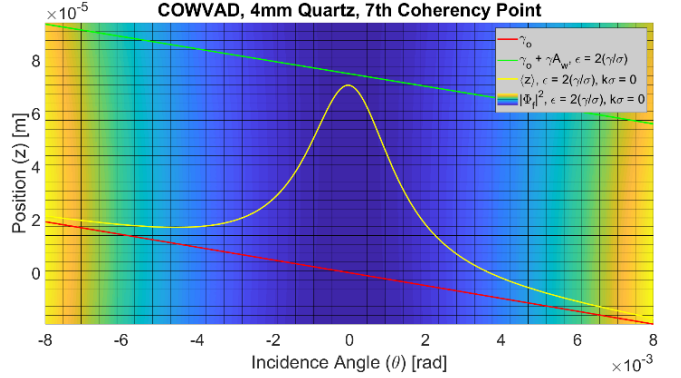

FIG. 7 Position expectation around the 7th coherency point. Near the coherency point, the position expectation (yellow line) deviates away from $\gamma_o$ (red line) and approaches the WVA prediction value (green line). The intensity of the beam is weakest at the coherency points and strongest midway between them.

## C. The Effect of a Momentum Boost

If a small relative momentum boost is present, i.e., the orthogonally polarized beams do not emerge from the crystal exactly parallel due to a slight misalignment of the optic axis, the behavior of the pointer around a coherency point may be changed drastically. In (10), it can be seen that this occurs when the second term dominates, which results in $\langle z \rangle$ taking an anti-symmetric form centered on the coherency point as seen in (12) and FIG. 8 rather than the symmetric form seen in FIG. 7. The first term can be made zero by the post-selection parameter while the second can be made zero by the phase-shift. Therefore, the pure WVA regime is defined as $k\sigma = 0$ and $\phi = 0$, where $\gamma/\sigma \ll \epsilon \ll 1$, and the pure inverse-WVA regime is defined by $\gamma = 0$ and $\epsilon = 0$, where $\phi \ll k\sigma \ll 1$. Both regimes demonstrate high sensitivity to incidence angle, but the inverse-WVA regime presents a form that may be easier to work with from the perspective of engineering practical applications. Note that for a birefringent crystal, $\gamma \approx 0$ only when $\theta = 0$.

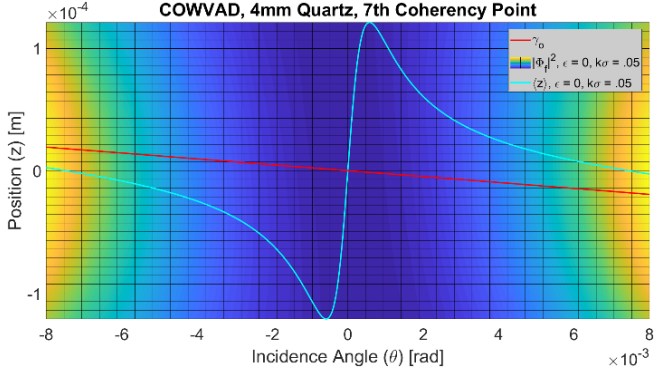

FIG. 8  Position expectation near the 7th coherency point in inverse-WVA regime, where $k\sigma > 0$ and $\epsilon \ll k\sigma$. In this regime, the curve is anti-symmetric, crossing $\gamma_o$ at the coherency point.

## VII. EXPERIMENT

### A. Setup

A photon beam is launched from a 633-nm Helium-Neon Laser (Newport R-32734) into free space. The photons traverse the preselection linear polarizer, which is oriented with its transmission axis at $\pi/4$ rad to the y-z-plane (local horizontal). Next, the preselected photons refract through a 4 mm birefringent crystal (Newlight Photonics BIR1040) oriented with its optic axis parallel to the x-axis (local vertical) such that $k\sigma \approx 0$. The crystal is held by a piezo controlled picomotor mount (Newport 8822) via a 1- to 2-inch optic adapter (ThorLabs AD2T) and is rotated about the z-axis to form an angle of incidence with the photon path. After refraction through the crystal, the photons traverse the post-selection polarizer, which is mounted in a stepper motor rotation mount (ThorLabs K10CR1), and oriented with its transmission axis at an angle of $(\epsilon - \pi/4)$ rad to the y-z-plane. Finally, the photons are detected by a CCD camera (Newport LBP2-HR-VIS2). All of the components are mounted in a straight line on an optical table with a distance of ~.5 m from laser to detector.

### B. Procedure

When the pre- and post-selection polarizers are crossed, coherency points are found by rotating the crystal about the z-axis to the angles where the beam intensity is minimized. At these points, a symmetric dual-mode Gaussian beam profile is produced at the detector. The symmetry of this profile is used to fine-tune $\epsilon$ and $\theta$. In this configuration, $\langle z \rangle = \gamma_o$ according to (10); therefore, a measurement of $\langle z \rangle$ is taken at each coherency point to establish a reference point $(\theta, z, \epsilon)$ from which all subsequent measurements are relative. The parameter space is navigated using the stepper motor rotation mount to precisely vary $\epsilon$ and the piezo picomotor mount to precisely vary $\theta$. Measurements of $\langle z \rangle$ are taken via the CCD and processed using the LPB2-HR-VIS2 software interface, which provides the mean and standard deviation of a user-defined number of camera frames. In this report, all measurements consist of data from 500 frames.

### C. Experimental Results

The lab data are displayed in FIG. 9 and FIG. 10. Though not originally intended, the curve traced by the measured data points for $\epsilon = 0$ indicates the presence of a momentum boost, implying that the optic axis of the birefringent crystal was not set exactly perpendicular to the plane of incidence. A value of $k\sigma \sim .05$ in eqn. (12) produces a good fit with the data taken at $\epsilon = 0$. This corresponds to a beam divergence angle $k/k_o \sim 30$ μrad.

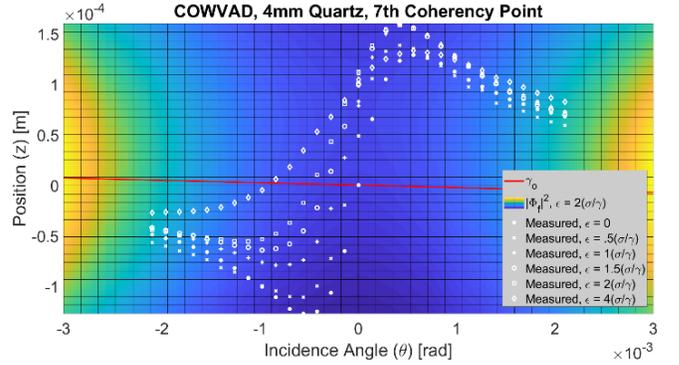

FIG. 9  Lab measurements of the position expectation values around the 7th coherency point for varying post-selection angles. The sensitivity of $\langle z \rangle$ to $\theta$ is maximized exactly at the coherency point when $\epsilon = 0$ rad.

The measured data indicates a promising trend for practical applications due to the steep slopes in $\langle z \rangle$ with respect to $\theta$ at each coherency point. The value of $\Delta\langle z \rangle/\Delta\theta$ appears to be maximized at each coherency point when the post-selection polarizer is orthogonal to the preselection polarizer ($\epsilon = 0$) as seen in FIG. 9.

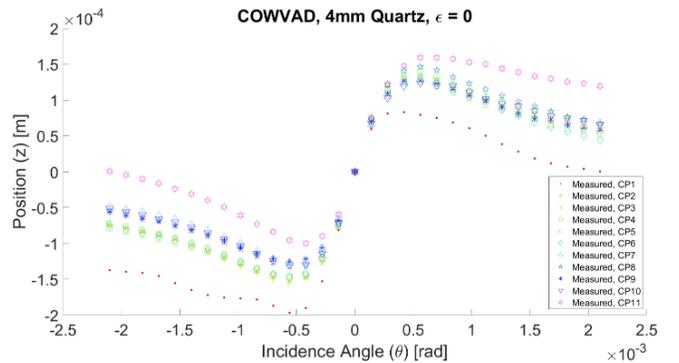

FIG. 10  Lab measurements of the position expectation values $\langle z \rangle$ of the first 11 coherency points with crossed polarizers. The reference measurement of each data series is aligned to the origin. The x- and y-axes are therefore displayed relative to each coherency point, $(\theta = \theta_{CP}, \langle z \rangle = \gamma_o(\theta_{CP}))$ where $\theta_{CP}$ are the angles of incidence for each coherency point.

The fact that $\Delta\langle z \rangle/\Delta\theta$ is maximized at each coherency point with crossed polarizers is a convenient result for practical applications. These are particularly easy values to locate due to indicators such as power minimization and the symmetric dual mode Gaussian beam profile. There is some variation in $\Delta\langle z\rangle/\Delta\theta$ from one coherency point to the next. For the 4 mm crystal, the $11^{th}$ coherency point ($\theta = .9955$ rad) is nearest to its neighboring points, which indicates that $\Delta\langle z\rangle/\Delta\theta$ might be largest at that point. However, the lab data (TABLE I and TABLE II), show the largest measured slope occurs instead at the *first* coherency point. It is worth noting that internal reflections and imperfections in the optics cause interference issues that distort the beam profile. This contributes to uncertainty in the lab data and is likely the reason for asymmetries and misalignments in FIG. 9 and FIG. 10.

TABLE I. Measured tilt sensitivity at the 7th coherency point for various post-selector values.

| Coher. Pt. # | $\epsilon\left(\frac{\gamma}{\sigma}\text{ rad}\right)$ | $\frac{\Delta\langle z\rangle}{\Delta\theta}\left(\frac{\text{m}}{\text{rad}}\right)$ |
|---|---|---|
| 7 | 0 | .49 |
| 7 | 1/2 | .47 |
| 7 | 1 | .37 |
| 7 | 3/2 | .32 |
| 7 | 2 | .27 |
| 7 | 4 | .13 |

TABLE II. Measured tilt sensitivity for crossed polarizers for various coherency points.

| Coher. Pt. # | $\epsilon\left(\frac{\gamma}{\sigma}\text{ rad}\right)$ | $\frac{\Delta\langle z\rangle}{\Delta\theta}\left(\frac{\text{m}}{\text{rad}}\right)$ |
|---|---|---|
| 1 | 0 | .58 |
| 2 | 0 | .55 |
| 3 | 0 | .52 |
| 4 | 0 | .54 |
| 5 | 0 | .54 |
| 6 | 0 | .54 |
| 7 | 0 | .49 |
| 8 | 0 | .52 |
| 9 | 0 | .49 |
| 10 | 0 | .51 |
| 11 | 0 | .53 |

## VIII. CONCLUSION

We have investigated the breakdown of the WVA effect in the COWVAD architecture with respect to angle of incidence of the laser on the birefringent crystal. This was done in order to determine the feasibility of using the crystal tilt as the sensitive component in a conceptual WVA-enhanced CVG.

We have expanded the theory describing WVA in the COWVAD to include the dependence of the observable pointer on the angle of incidence. This new model describes the behavior of the pointer over the full range of incidence angles where previous studies only considered the discrete set of coherency-point angles. The model predicts rapid pointer shifts for small angular deflections near coherency points, which equates to amplified response to crystal tilt. Additionally, we have expanded the theory to account for the presence of a momentum boost, which describes configurations in which the orthogonally polarized beams exit the crystal on paths that are not exactly parallel. We found that the presence of a slight momentum boost drastically changes the behavior of the pointer in the tilt sensitive region. Choosing pre- and post-selected polarization states that are exactly orthogonal maximizes the effect as the result of the COWVAD being placed into an inverse-WVA configuration. This configuration opens new paths to practical tilt sensitive applications.

In addition to the already well-known coherency points, we found that WVA can be made to occur at angles where the orthogonally polarized beam components emerge from the birefringent crystal $\pi/2$ rad out of phase (Appendix B). WVA at these anti-coherency points is achieved by setting the pre- and post-selection polarizers to be nearly *aligned*. This was predicted by the mathematical model and confirmed in the lab. The presence of additional amplification points is beneficial for flexibility in practical applications.

We performed a laboratory experiment measuring the tilt sensitivity of the COWVAD. The shape of the resulting data is well matched by the momentum-boost model. The maximum sensitivity we measured was ~580 mm/rad. This is 390 times greater than the classical $\theta$-sensitivity at the first coherency point without amplification where $|d\gamma_o/d\theta| = 1.5$ mm/rad. Note that this result does not represent the best possible performance of the device, but serves to confirm the mathematical model. Also note that the amplification in sensitivity is independent of the distance between optical components; thus the overall length of the device can be shortened arbitrarily.

These results are extremely promising for the overarching goal of realizing a WVA-enhanced CVG. The mathematical model developed in this report reveals the interplay between the key parameters and how they can be manipulated to achieve even better results. This model is confirmed by the experimental data. Further investigation is warranted to analyze the noise characteristics of this device and investigate configurations that will maximize the sensitivity of the pointer to the tilt of the crystal.

### AKNOWLEDGEMENTS

The authors would like to thank Ms. Jennifer Clift and Dr. Jeffrey Solka for their support of weak-value research over the years. Without their dedication and advocacy, this research would not have been possible. We also want to thank Dr. Parks and Scott Spence, not only for their foundational research in weak-value theory that laid the groundwork for these new developments, but also for their mentorship over the years. This work was supported by the Navy Innovative Science and Engineering program and the In-house Laboratory Independent Research program.

## APPENDIX A

### A. Canonical, Weak-Value Treatment

The surprising prediction of weak-value theory is that a measurement of a quantum mechanical observable can yield a value that is far outside its eigenvalue spectral range [1]. This was demonstrated in [16] using the COWVAD specifically at coherency points ($\cos\phi = 1$) when the beams exit the birefringent crystal exactly parallel [16]. In these cases, (5) can be simplified and expanded to

$$\widehat{U} \approx e^{-\frac{i}{\hbar}\gamma_o \hat{p}}\left(1 - \frac{i}{\hbar}\gamma\hat{p}\otimes\hat{A}\right). \tag{A1}$$

where expansion of the exponential term is a valid approximation when the photon-crystal interaction is weak ($\gamma/\sigma \ll 1$). Next, by applying the approximate form of $\widehat{U}$ to the preselected qubit state $|\psi_i\rangle$, projecting the resulting state onto the post-selected qubit state $|\psi_f\rangle$, and reapproximating the $\gamma$ term as an exponential, the equation can be written in a revealing form

$$\langle\psi_f|\widehat{U}|\psi_i\rangle \approx \langle\psi_f|\psi_i\rangle e^{-\frac{i}{\hbar}(\gamma_o + \gamma A_w)\hat{p}} \tag{A2}$$

where the weak value is defined as

$$A_w = \frac{\langle\psi_f|\hat{A}|\psi_i\rangle}{\langle\psi_f|\psi_i\rangle} = \cot\epsilon. \tag{A3}$$

Here we have made the substitution $\beta = \epsilon - \pi/4$ such that $\epsilon$ is a measure of the deviation of the post-selection state away from orthogonality with the preselection state. The approximations made in calculating (A2) lead to a set of conditions that must be met in order for the equation to be valid

$$\frac{\gamma}{\sigma} \ll \epsilon \ll 1. \tag{A4}$$

These are the weak-value requirements for the COWVAD. The final state of the pointer takes the form

$$|\Phi_f\rangle \approx \sin(\epsilon)\, e^{-\frac{i}{\hbar}(\gamma_o + \gamma \cot(\epsilon))\hat{p}}|\Phi_i\rangle. \tag{A5}$$

It can be seen in (A5) that when $\epsilon$ is made to be very small by choosing the post-selected qubit state to be nearly orthogonal to the preselected state, the weak value becomes very large, effectively amplifying the spatial translation of the pointer by a factor of $\cot\epsilon$ at the cost of attenuating the overall amplitude by a factor of $\sin\epsilon$.

## APPENDIX B

### A. Probability of Post-Selection

The probability that a photon will survive post-selection is given by the magnitude of the post-unitary state projected onto the post-selected state, which is simply the normalization factor of the final state wave function

$$\begin{aligned}P &= \int_{-\infty}^{\infty}|\langle z|\Phi_f\rangle|^2 dz \\ &= \frac{1}{2}\left[1 - \cos(2\epsilon)\cos\phi\, e^{-\frac{(k\sigma)^2 + (\gamma/\sigma)^2}{2}}\right].\end{aligned} \tag{B1}$$

Because the interaction strength variables $\gamma$ and $\phi$ are functions of $\theta$, the probability of post-selection $P$ is also $\theta$ dependent (FIG. 11).

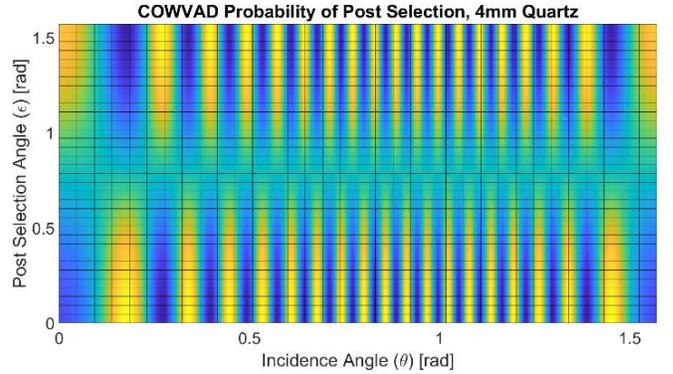

FIG. 11  Probability of post-selection dependence on incidence angle and post-selector angle.

Along the bottom of FIG. 11 ($\epsilon \ll 1$), the dark bands represent the coherency points where the beams are approximately in phase and WVA occurs. Interestingly, WVA also occurs at another set of dark bands produced where $\epsilon \approx \pi/2$ (pre- and post-selected states nearly parallel) at $\theta$-values for which the two photon paths are nearly out of phase ($\cos\phi \approx -1$). Classically, this can be understood by the fact that $\phi$ describes the generally elliptical polarization state of the light exiting the crystal. When the overlapping beams are perfectly in phase, photons leaving the crystal have the same polarization as photons entering the crystal. However, when the overlapping beams are perfectly out of phase, the crystal has rotated the polarization state of the photons such that the exiting photons are polarized perpendicular to those entering the crystal. Therefore, a post-selector nearly aligned with the preselector will select a polarization nearly orthogonal to that exiting the crystal.

## B. Weak Values at Anti-Coherency Points
## (An Interesting Side-Note)

As suggested in FIG. 11, the weak value is defined when the pre- and post-selection polarizers are nearly aligned. At these alternate coherency points, the phase-shift produced in the crystal results in a $\pi/2$ rotation of the polarization of the beam exiting the crystal. The final state of the system then takes the form

$$|\psi_f\rangle = \cos\beta\,|H\rangle - \sin\beta\,|V\rangle \quad (B2)$$

which produces the weak value

$$\frac{\langle\psi_f|\hat{A}|\psi_i\rangle}{\langle\psi_f|\psi_i\rangle} = -\cot\epsilon \quad (B3)$$

where this time we have made the substitution $\beta = \epsilon + \pi/4$ and $\epsilon$ is some small angle deviation from *alignment* between the pre- and post-selection polarizers. This is further verified by the position expectation value, which takes the form

$$\langle z \rangle = \gamma_o - \frac{\gamma\sin(2\epsilon) + k\sigma^2\cos(2\epsilon)\sin\phi\, e^{-\frac{(k\sigma)^2+(\gamma/\sigma)^2}{2}}}{1+\cos(2\epsilon)\cos\phi\, e^{-\frac{(k\sigma)^2+(\gamma/\sigma)^2}{2}}} \quad (B4)$$

$$\approx \gamma_o - \gamma\cot\epsilon$$

where the approximation on the right is valid when $k\sigma = 0$, $\cos\phi \approx -1$ and $e^{-\gamma^2/2\sigma^2} \approx 1$.